\documentclass[12pt,preprint]{aastex}
\newcounter{address}
\newcommand{\latin}[1]{{#1}}
\newcommand{\ie}{\latin{i.e.}}
\newcommand{\eg}{\latin{e.g.}}
\newcommand{\cf}{\latin{c.f.}}

\begin{document}

\title{The overdensities of galaxy environments as a function of
       luminosity and color}
\author{
  David~W.~Hogg\altaffilmark{\ref{NYU},\ref{email}},
  Michael~R.~Blanton\altaffilmark{\ref{NYU}},
  Daniel~J.~Eisenstein\altaffilmark{\ref{Steward}},
  James~E.~Gunn\altaffilmark{\ref{Princeton}},
  David~J.~Schlegel\altaffilmark{\ref{Princeton}},
  Idit~Zehavi\altaffilmark{\ref{FNAL},\ref{Chicago}},
  Neta~A.~Bahcall\altaffilmark{\ref{Princeton}},
  Jon~Brinkmann\altaffilmark{\ref{APO}},
  Istvan~Csabai\altaffilmark{\ref{Eotvos},\ref{JHU}},
  Donald~P.~Schneider\altaffilmark{\ref{PSU}},
  David~H.~Weinberg\altaffilmark{\ref{OSU}},
  Donald~G.~York\altaffilmark{\ref{Chicago}}
}

\setcounter{address}{1}
\altaffiltext{\theaddress}{\stepcounter{address}\label{NYU}
Center for Cosmology and Particle Physics, Department of Physics, New
York University, 4 Washington Place, New York, NY 10003}
\altaffiltext{\theaddress}{\stepcounter{address}\label{email}
\texttt{david.hogg@nyu.edu}}
\altaffiltext{\theaddress}{\stepcounter{address}\label{Steward}
Steward Observatory, 933 N. Cherry Ave., Tucson, AZ 85721}
\altaffiltext{\theaddress}{\stepcounter{address}\label{Princeton}
Princeton University Observatory, Princeton, NJ 08544}
\altaffiltext{\theaddress}{\stepcounter{address}\label{FNAL}
Fermi National Accelerator Laboratory, PO Box 500, Batavia, IL 60510}
\altaffiltext{\theaddress}{\stepcounter{address}\label{Chicago}
Astronomy and Astrophysics Department, University of Chicago, Chicago, IL 60637}
\altaffiltext{\theaddress}{\stepcounter{address}\label{APO}
Apache Point Observatory, 2001 Apache Point Road,
P.O. Box 59, Sunspot, NM 88349-0059}
\altaffiltext{\theaddress}{\stepcounter{address}\label{Eotvos}
Department of Physics, E\"{o}tv\"{o}s University,
Budapest, Pf.\ 32, Hungary, H-1518}
\altaffiltext{\theaddress}{\stepcounter{address}\label{JHU}
Department of Physics and Astronomy, The Johns Hopkins University,
Baltimore, MD 21218}
\altaffiltext{\theaddress}{\stepcounter{address}\label{PSU}
Department of Astronomy and Astrophysics,
The Pennsylvania State University, University Park, PA 16802}
\altaffiltext{\theaddress}{\stepcounter{address}\label{OSU}
Department of Astronomy, Ohio State University, Columbus, OH 43210}

\begin{abstract}
We study the mean environments of galaxies in the Sloan Digital Sky
Survey as a function of rest-frame luminosity and color.
Overdensities in galaxy number are estimated in
$8\,h^{-1}~\mathrm{Mpc}$ and $1\,h^{-1}~\mathrm{Mpc}$ spheres centered
on $125,000$ galaxies taken from the SDSS spectroscopic sample.  We
find that, at constant color, overdensity is independent of luminosity
for galaxies with the blue colors of spirals.  This suggests that, at
fixed star-formation history, spiral-galaxy mass is a very weak
function of environment.  Overdensity does depend on luminosity for
galaxies with the red colors of early types; both low-luminosity and
high-luminosity red galaxies are found to be in highly overdense
regions.
\end{abstract}


\section{Introduction}

Elliptical and lenticular galaxies are over-represented in massive
nearby galaxy clusters relative to the field \citep{dressler80a,
postman84a}.  Elliptical galaxies also tend to be redder, more
luminous, more metal-rich, more gas-poor, and older in stellar
population than spirals and irregulars \citep[\eg,][]{tammann79a,
kormendy89a, roberts94a}.  Indeed, it has also been found that color,
luminosity, surface-brightness, gas content, stellar population age,
and star-formation rate are all correlated with the overdensity of the
galaxy environment \citep[\eg,][]{kennicutt83a, balogh01a, martinez02a,
lewis02a, blanton02d, gomez02q}.  Along the same lines, studies of the
clustering of galaxies have found different clustering amplitudes for
galaxies of different types, colors, and luminosities
\citep[\eg,][]{davis76a, mo94a, park94a, norberg02a, zehavi02a}.  What
is not understood is which of the relationships with environment are
causal and which are just a by-product of other, more fundamental
correlations.

Conventional cosmological theories posit that galaxies reside inside
dark matter concentrations or ``halos'' that grow from small random
fluctuations at early times.  The most overdense fluctuations will
collapse first; in a gaussian random field, these preferentially
reside within overdensities on larger scales, implying a correlation
of halo mass with environment is expected \citep[\eg,][]{mo96a,
lemson99a}.  The relationships between the properties of a halo and
the properties of the galaxy or galaxies it contains are not fully
understood, so the conventional theories do not currently make strong
predictions for the dependence of galaxy number overdensity on
observable galaxy properties, although most studies suggest that
luminosity and color will be related to overdensity
\citep[eg,][]{kauffmann97a, kauffmann99a, benson00a}.  Whether or not
this dark-matter halo picture ends up being useful or correct, and
whatever turn out to be the important physical processes for making
galaxies, the investigations started in this \textsl{Letter} will
place important constraints on galaxy formation and evolution.

The Sloan Digital Sky Survey (SDSS) is the best available data set for
investigation of these relationships because of its sample size, high
signal-to-noise imaging, sky coverage, and complete spectroscopy
\citep[\eg,][]{york00a}.  Indeed, the SDSS has already made some
relevant measurements, including the dependence of clustering on
luminosity and color \citep{zehavi02a}, the star-formation as a
function of environment \citep{gomez02q}, and the mean red-galaxy
spectrum as a function of environment \citep{eisenstein02a}.  In this
\textsl{Letter,} we investigate the mean galaxy number overdensities
around galaxies of different colors and luminosities.

In what follows, a cosmological world model with
$(\Omega_\mathrm{M},\Omega_\mathrm{\Lambda})=(0.3,0.7)$ is adopted,
and the Hubble constant is parameterized
$H_0=100\,h~\mathrm{km\,s^{-1}\,Mpc^{-1}}$, for the purposes of
calculating distances and volumes \citep[\eg,][]{hogg99cosm}.

\section{Data sample}

The SDSS is taking $ugriz$ CCD imaging of $10^4~\mathrm{deg^2}$ of the
Northern Galactic sky, and, from that imaging, selecting $10^6$
targets for spectroscopy, most of them galaxies with
$r<17.77~\mathrm{mag}$ \citep[\eg,][]{gunn98a,york00a,stoughton02a}.

All the data processing: astrometry \citep{pier02a}; source
identification, deblending and photometry \citep{lupton01a};
calibration \citep{fukugita96a,smith02a}; spectroscopic target
selection \citep{eisenstein01a,strauss02a,richards02a}; spectroscopic
fiber placement \citep{blanton02a}; and spectroscopic data reduction
are performed with automated SDSS software.

Redshifts are measured on the reduced spectra by an automated system,
which models each galaxy spectrum as a linear combination of stellar
populations (Schlegel, in preparation).

The sample is statistically complete, with small incompletenesses
coming primarily from (1) galaxies missed because of mechanical
spectrograph constraints \citep[6~percent;][]{blanton02a}, which does
lead to a slight under-representation of high-density regions, and (2)
spectra in which the redshift is either incorrect or impossible to
determine ($<1$~percent).  In addition, there are some galaxies ($\sim
1$~percent) blotted out by bright Galactic stars, but this
incompleteness should be uncorrelated with galaxy properties.

For the purposes of computing large-scale structure statistics, we
have assembled a subsample of SDSS galaxies known as the NYU LSS
\texttt{sample10}.  This subsample includes not only properties of the
galaxies but also of survey selection function variations and angular
coverage.  For each galaxy in \texttt{sample10}, the sample includes a
computed volume $V_\mathrm{max}$ representing the total volume of the
Universe (in $h^{-3}~\mathrm{Mpc^3}$) in which the galaxy could have
resided and still made it into the sample.  The calculation of these
volumes is described elsewhere \citep{blanton02d}.

Galaxy luminosities and colors \citep[measured by the standard SDSS
petrosian technique;][]{petrosian76a} are computed in fixed
bandpasses, using Galactic extinction corrections \citep{schlegel98a}
and $K$ corrections \citep[computed with \texttt{kcorrect
v1\_11};][]{blanton02b}.  They are $K$ corrected not to the redshift
$z=0$ observed bandpasses but to bluer bandpasses $^{0.1}g$, $^{0.1}r$
and $^{0.1}i$ ``made'' by shifting the SDSS $g$, $r$, and $i$
bandpasses to shorter wavelengths by a factor of 1.1
\citep[\cf,][]{blanton02b, blanton02d}.  This means that galaxies at
redshift $z=0.1$ (typical of the SDSS sample used here) have trivial
$K$ corrections.

The sample of galaxies used here was selected to have apparent
magnitude in the range $14.5<r<17.77~\mathrm{mag}$, redshift in the
range $0.05<z<0.22$, and fixed-frame absolute magnitude in the range
$M_{^{0.1}i}>-24.0~\mathrm{mag}$.  These cuts left 124884 galaxies.

Overdensity estimators on two different length scales are used.  On
the $8\,h^{-1}~\mathrm{Mpc}$ scale, the estimate of environment
overdensity $\delta_8$ is based on the SDSS spectroscopic sample.  It
is a measure of the three-dimensional redshift-angle space number
density excess around each galaxy.  The comoving transverse distances
and comoving line-of-sight distances \citep[\eg,][]{hogg99cosm} are
computed between each spectroscopic galaxy and its neighboring
spectroscopic galaxies (not attempting to correct for peculiar
velocities).  Neighbors within an $8\,h^{-1}~\mathrm{Mpc}$ comoving
sphere in this space are counted; the result is divided by the
prediction made from the galaxy luminosity function
\citep{blanton02c}, and unity is subtracted to produce the overdensity
estimate $\delta_8$.  A galaxy in an environment with the cosmic mean
density has $\delta_8=0$.  Although the sample used to infer
$\delta_8$ is flux-limited and not volume-limited, the resulting
overdensity estimates have been shown to be redshift-independent in
the median \citep{blanton02d}.  If the spatial correlation function
$\xi(r)$ has the form $r^{-\gamma}$, then the mean overdensity around
galaxies $\left<\delta_8\right>$ will be
$(3/[3-\gamma])\,\xi(8\,h^{-1}~\mathrm{Mpc})$.  Direct comparison of
the $\left<\delta_8\right>$ and $\xi(r)$ requires a correction for the
presence of redshift distortions due to infall and ``fingers of God''.

On the $1\,h^{-1}~\mathrm{Mpc}$ scale, the estimate of environment
overdensity $\delta_1$ is a deprojected angular correlation function.
Around each spectroscopic target galaxy, galaxies are counted in the
SDSS imaging in the magnitude range corresponding to
$M^\ast\pm1~\mathrm{mag}$ (passively-evolved and $K$-corrected as for
an early-type galaxy) and within $5h^{-1}~\mathrm{Mpc}$
\citep[transverse, proper; \eg,][]{hogg99cosm} at the spectroscopic
galaxy redshift.  The count is weighted so as to recover the estimated
overdensity averaged over a \emph{spherical} three-dimensional
Gaussian window $e^{-r^2/2a^2}$ with a radius of
$a=1h^{-1}~\mathrm{Mpc}$ (proper).  Details of the weighting and the
method for correcting for the survey mask are given elsewhere
\citep{eisenstein02b}.  The results do not depend on an assumed model
of the correlation function but do depend inversely on the
normalization of the luminosity function at the redshift in question.
One advantage of this method is that the density can be estimated with
a volume-limited and yet reasonably dense set of galaxies, even at the
furthest reaches of the spectroscopic catalog.  Another advantage is
that the estimator is not affected by redshift distortions.  If the
spatial correlation function $\xi(r)$ has the form $r^{-\gamma}$, then
the mean overdensity around galaxies $\left<\delta_1\right>$ is
$(2/\sqrt{\pi})\,\Gamma([3-\gamma]/2)\,\xi(1\,h^{-1}~\mathrm{Mpc})$.

The two overdensity estimators $\delta_1$ and $\delta_8$ are very
different; the deprojection technique is not possible on the large
scale because of survey geometry constraints, and the redshift-angle
sphere technique is not possible on the small scale because of
peculiar velocities.  Of course, that different estimators with
different systematics are used on the two different scales makes their
inter-comparison informative.

\section{Results}

\begin{figure}
\noindent\includegraphics{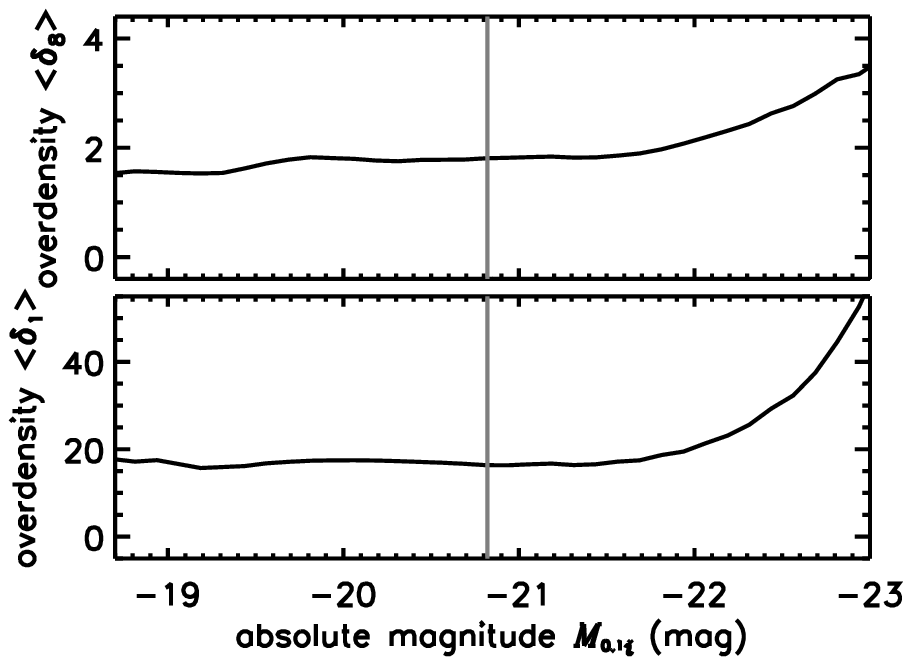}\includegraphics{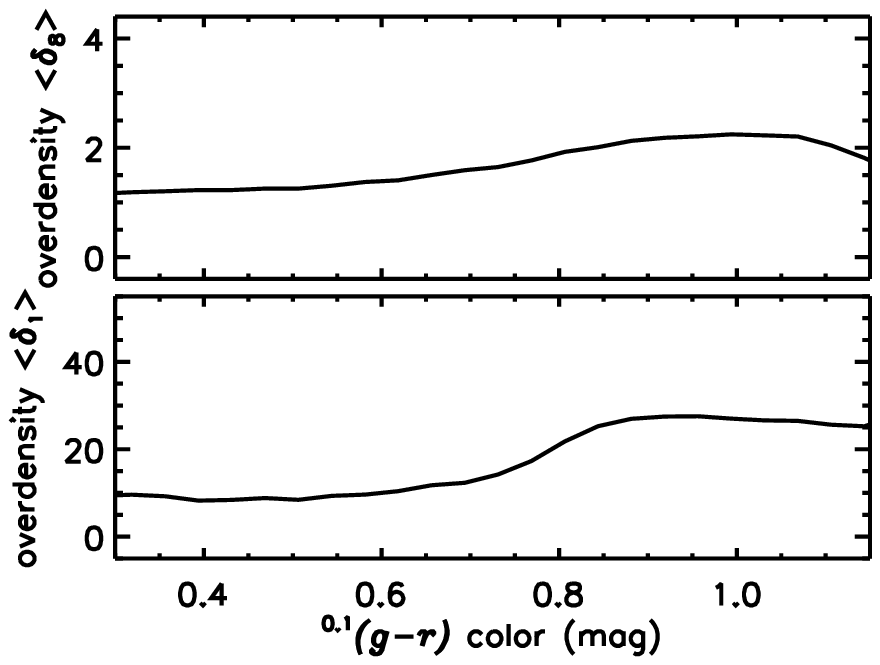}
\caption[]{The $1/V_\mathrm{max}$-weighted mean galaxy environment
overdensities $\left<\delta_8\right>$ and $\left<\delta_1\right>$ in
$8\,h^{-1}~\mathrm{Mpc}$ and $1\,h^{-1}~\mathrm{Mpc}$ spheres (see
text) computed in a sliding box of luminosity width $\Delta
M_{^{0.1}i}=0.5~\mathrm{mag}$ (absolute magnitude plots) or in a
sliding box of color width $\Delta ^{0.1}(g-r)=0.15~\mathrm{mag}$
(color plots).  A vertical line shows the characteristic luminosity
$L^\ast$ \citep{blanton02c}.\label{fig:1d}}
\end{figure}

Fig~\ref{fig:1d} shows the mean dependencies of overdensity on
luminosity and color separately.  Both show monotonic trends, with
redder and more luminous galaxies, on average, in higher density
regions.  In each panel the mean is weighted by the inverse selection
volumes $1/V_\mathrm{max}$.

Panel~\textsl{(a)} of Fig~\ref{fig:2d} shows the distribution of
galaxies in the sample in the luminosity-color plane.  The ``red
sequence'' of old stellar populations, \ie, early-type galaxies, is
visible.  We will refer to galaxies on this sequence as ``red'' and
those bluer as ``blue.''  The mode of the luminosity distribution is
more luminous than $L^\ast$.  This is due to the classical Malmquist
bias: more luminous galaxies are visible over a much larger volume and
are therefore over-represented in the sample.  Taking means weighted
by $1/V_\mathrm{max}$ compensates for this bias.

Panels~\textsl{(b)} and \textsl{(c)} of Fig~\ref{fig:2d} show the
weighted mean environment overdensities $\left<\delta_8\right>$ and
$\left<\delta_1\right>$ computed in a sliding box of luminosity width
$\Delta M_{^{0.1}i}=0.5~\mathrm{mag}$ and color width $\Delta
{}^{0.1}(g-r)=0.15~\mathrm{mag}$, at all colors and magnitudes at
which the sliding box contains at least 200 galaxies.  The trends of
the two different density estimators are remarkably similar:

(1)~Galaxies of every color and luminosity are, on average, in
   overdense regions ($\delta > 0$).  There are no unclustered or void
   populations that can be isolated by color and luminosity, in the
   color and luminosity ranges considered here.

(2)~Very luminous ($L>3\,L^\ast$) red galaxies are, on average, in
   more overdense regions than the average galaxy.  This is not
   surprising, given that many clusters contain extremely luminous
   galaxies \citep[\eg,][]{beers83a}.

(3)~Faint ($L<(1/3)\,L^\ast$) red galaxies are also, on average, in
   more overdense regions than the average galaxy.  This may be
   related to the observation that the luminosity function appears
   more dwarf-rich in clusters than in the field
   \citep[\eg,][]{driver94b, bernstein95a, depropris95a, trentham02a}.

(4)~The mean overdensity of blue galaxies is independent of
   luminosity and increases with color, at least at luminosities
   $L<3\,L^\ast$.  That overdensity is related to color reflects the
   color--morphology and density--morphology relations
   \citep{dressler80a, postman84a, roberts94a}.

\section{Discussion}

We have shown that blue galaxies, \ie, galaxies bluer than the red
sequence of old stellar populations, exhibit no correlation between
their luminosities and the overdensity of their environments, at fixed
color.  The class of blue galaxies includes most spirals and
irregulars.

\begin{figure}
\noindent\includegraphics{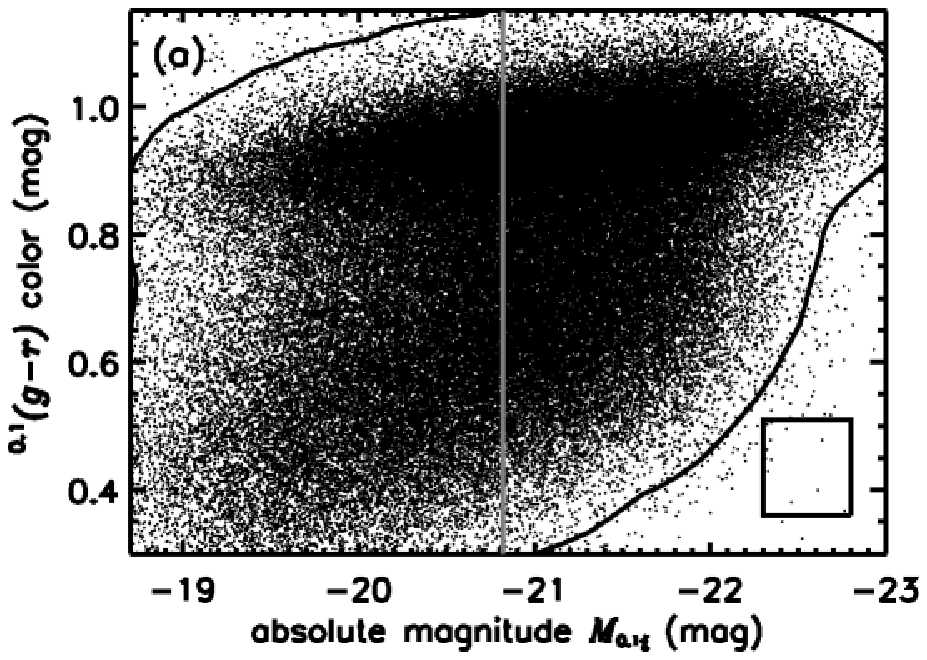}\\[1ex]
\includegraphics{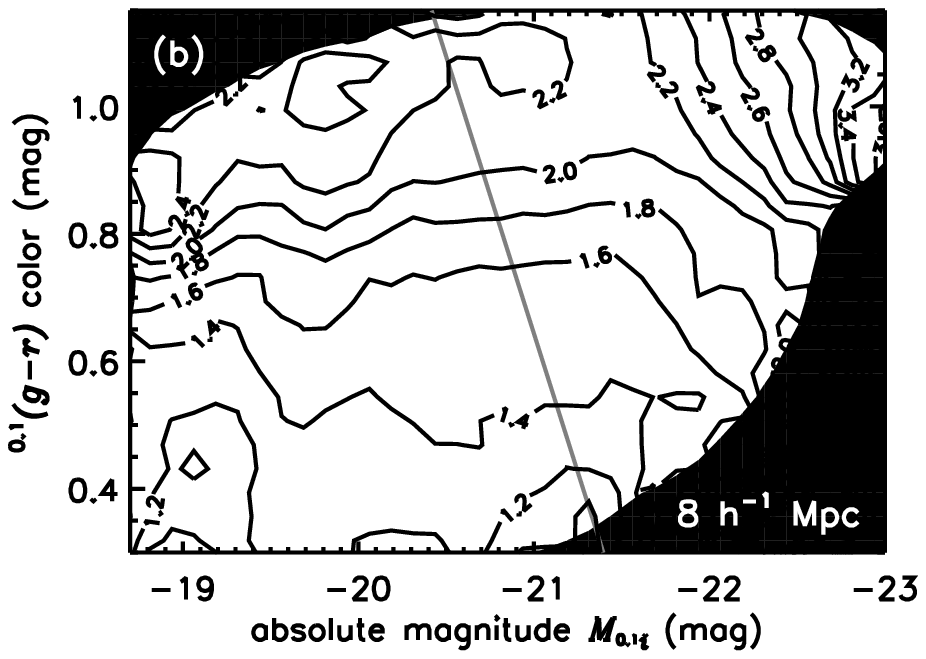}
\includegraphics{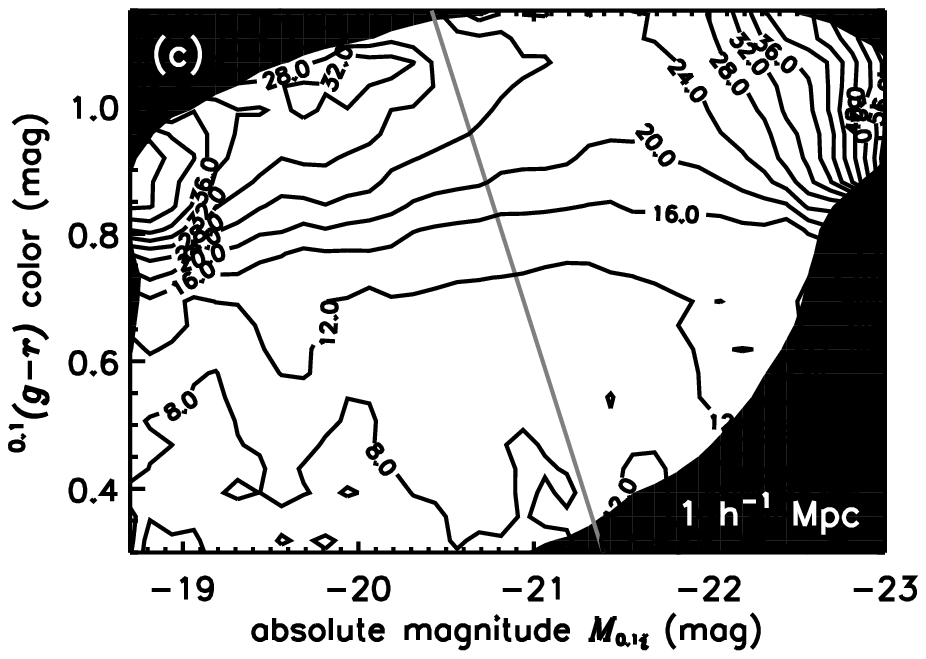}
\caption[]{Panel~\textsl{(a)} shows the luminosity and color of each
galaxy in the sample.  A vertical grey line shows the characteristic
luminosity $L^\ast$ \citep{blanton02c}.  Panel~\textsl{(b)} shows the
$1/V_\mathrm{max}$-weighted mean galaxy environment overdensity
$\left<\delta_8\right>$ in $8\,h^{-1}~\mathrm{Mpc}$ spheres (see text)
computed in a sliding, rectangular box of luminosity width $\Delta
M_{^{0.1}i}=0.5~\mathrm{mag}$ and color height $\Delta
^{0.1}(g-r)=0.15~\mathrm{mag}$.  The box size and shape is shown on
the lower right corner of panel~\textsl{(a)}.  The grey line
indicates, roughly, a locus of constant total stellar mass
\citep{bell01b}.  Panel~\textsl{(c)} shows the weighted mean
overdensity $\left<\delta_1\right>$ in $1\,h^{-1}~\mathrm{Mpc}$
spheres (see text).  Note that different kinds of overdensity
estimators are used for the two different length scales.  The mean
environment overdensity is not shown shown at colors and luminosities
at which there are $<200$ galaxies inside the box; this region is
indicated with a border on panel~\textsl{(a)}.\label{fig:2d}}
\end{figure}

On Fig~\ref{fig:2d} a line of (roughly) constant stellar mass is
shown, derived by converting relations derived in the $B$ and $R$
bands \citep{bell01b}; the conversion is straightforward because $B$
and $R$ are very similar to $^{0.1}g$ and $^{0.1}i$.  The line of
constant mass is close to a line of constant luminosity.  Similarly, a
line of constant star-formation history is close to a line of constant
color, with a small slope arising from the luminosity--metallicity
relation \citep[\eg,][]{vila-costas92a, zaritsky94a, ryder95a}.
Fig~\ref{fig:2d} shows that, for spiral galaxies, mean overdensity is
monotonically related to star-formation history.  That there is a
strong relationship between star-formation history and environment has
been suggested before, both theoretically \citep[eg,][]{kauffmann99a,
benson00a} and observationally \citep{balogh01a, gomez02q}.  What is
more remarkable is that, for galaxies with the colors of spirals, at
fixed star-formation history, overdensity does not appear to depend on
stellar mass at all.

The reddest galaxies show strong trends in overdensity at both the
luminous and faint extremes.  The former trend indicates that the very
most luminous galaxies lie in the very largest overdensities; probably
this is related to the fact that giant galaxies lie near the centers
of clusters.  The latter trend suggests that though high-density
regions are expected to be rich in dwarf galaxies, those dwarfs will,
by and large, be much redder than typical galaxies of their luminosity
(\citealt{norberg02a}; Zehavi et al, in preparation).  That red
galaxies have a minimum in their mean overdensity at luminosities near
$L^\ast$ could be partly due to the specific mixture of galaxy types
at those luminosities; there is a large overlap with the distribution
of spirals, and the red population contains a large number of Sa
galaxies at those luminosities \citep[\eg,][]{strateva01q}.

At one point it was thought that dwarf galaxies might ``fill the
voids,'' making them an underdense population in the mean.  Our
results rule this out, at least for dwarf galaxies selected by color
and luminosity in the ranges considered here.

\acknowledgments We thank Andreas Berlind, Douglas Finkbeiner, Jill
Knapp, Ravi Sheth, Roman Scoccimarro, Iskra Strateva, and Simon White
for useful discussions and software.  This research made use of the
NASA Astrophysics Data System.  MRB and DWH are partially supported by
NASA (grant NAG5-11669) and NSF (grant PHY-0101738). DJE is supported
by NSF (grant AST-0098577) and by an Alfred P. Sloan Research
Fellowship.

Funding for the creation and distribution of the SDSS has been
provided by the Alfred P. Sloan Foundation, the Participating
Institutions, the National Aeronautics and Space Administration, the
National Science Foundation, the U.S. Department of Energy, the
Japanese Monbukagakusho, and the Max Planck Society. The SDSS Web site
is {\tt http://www.sdss.org/}.

The SDSS is managed by the Astrophysical Research Consortium (ARC) for
the Participating Institutions. The Participating Institutions are The
University of Chicago, Fermilab, the Institute for Advanced Study, the
Japan Participation Group, The Johns Hopkins University, Los Alamos
National Laboratory, the Max-Planck-Institute for Astronomy (MPIA),
the Max-Planck-Institute for Astrophysics (MPA), New Mexico State
University, University of Pittsburgh, Princeton University, the United
States Naval Observatory, and the University of Washington.

\bibliographystyle{apj}
\bibliography{apj-jour,ccpp}

\end{document}